\documentclass{luxenote}

\usepackage[symbol]{footmisc}
\usepackage{textcomp}

\title{TARDIS Lab: Infrastructure for Characterization and Assembly of Highly Granular Silicon Calorimeter Modules for LUXE and Future Higgs Factories}

\author{Melissa Almanza-Soto\footnote[1]{Corresponding author melissa.almanza@ific.uv.es}}
\author{C\'esar Blanch}
\author{Shan Huang\footnote[1]{Corresponding author shan.huang@ific.uv.es}}
\author{Adri\'an Irles\footnote[1]{Corresponding author adrian.irles@ific.uv.es}}
\author{Carlos Orero}

\affil{IFIC, Universitat de Val\`encia and CSIC, C./ Catedr\'atico Jos\'e Beltr\'an 2, E-46980 Paterna, Spain}

\begin{document}

\maketitle

\begin{abstract}The TARDIS Lab has been established at IFIC (CSIC–UV) to develop infrastructure and expertise in high-granularity silicon calorimetry. The laboratory provides an ISO7 clean-room environment and dedicated instrumentation for silicon sensor characterization and ultra-thin calorimeter module assembly. It supports developments of silicon-based high-granularity calorimetry concepts pursued within CERN DRD Calo (DRD6) and the FCAL Collaboration. These developments serve not only as technological prototypes of the electromagnetic calorimeters and luminometers at future Higgs-factory proposals (FCC-ee, Linear Collider Facility at CERN, International Linear Collider, etc.) but are also essential for the strong-field QED program of the LUXE experiment. 
We describe the laboratory infrastructure, assembly workflows and DESY test-beam validation of fully assembled calorimeter modules. The assembled ECALp modules have an average thickness of approximately 760~\textmu m, while laboratory tests of 4352 channels identified connectivity problems in only 0.5\% of them. For the two SiWECAL modules assembled in 2025, connectivity problems were observed at the level of approximately $10^{-3}$ of the instrumented pads, with no mechanical failures or sensor delamination after transport and beam operation. We also present ongoing infrastructure upgrades and production milestones for 2026--2027.
\end{abstract}

\vfill
\begin{center}
\textit{Contribution to the International Workshop on Future Linear Colliders (LCWS 2025), 20-24 October 2025. Valencia, Spain (C25-10-20.1)}
\end{center} 
\clearpage


\section{Introduction}
\label{sec:intro}

Proposed electron--positron facilities such as the ILC (Japan)~\cite{Behnke:2013xla}, CEPC (China)~\cite{CEPCStudyGroup:2023quu}, FCC-ee (CERN)~\cite{FCC:2025lpp} or the LCF (CERN)~\cite{LinearCollider:2025lya} aim at precision measurements of the Higgs boson, the top quark and electroweak observables.

In pursuit of the most precise measurement systems, highly granular silicon--tungsten sampling calorimeters provide fine shower imaging, calibration stability, compact mechanical integration and enhanced Particle Flow reconstruction capabilities~\cite{Thomson:2009rp,Sefkow:2015hna}.
Luminosity measurement systems play a crucial role in the design of the detector systems. In particular, forward luminometers determine the integrated luminosity through precisely calculable QED processes, such as small-angle Bhabha scattering. Achieving the targeted precision requires detectors that are extremely compact, highly granular and mechanically stable, with micrometric control of the geometry in order to reconstruct the energy and angle of electromagnetic showers with high accuracy.

Two major collaborations established the conceptual and technological foundations of high-granularity silicon-based calorimetry: CALICE~\cite{CALICE} developed the highly integrated SiWECAL~\cite{Kawagoe:2019dzh} concept for barrel calorimetry, while FCAL~\cite{FCAL} developed ultra-compact forward calorimeters such as LumiCal and BeamCal~\cite{LumiCal,BeamCal}. Both collaborations concluded their initial activity phases and their developments are now integrated within CERN DRD Calo~\cite{DRDCalo} and the re-established FCAL Collaboration.

Today, instrumentation R\&D is coordinated within the CERN DRD Calo (DRD6) program, which focuses on sensor development, electronics, prototyping and test-beam validation. In parallel, the FCAL Collaboration provides the experiment-oriented framework, linking hardware developments with physics optimization and luminosity studies.
Moreover, both technological lines are currently being realized as detectors for the LUXE experiment:

\begin{itemize}
\item the ECALe, electromagnetic calorimeters for electrons (based on the SiWECAL concept),
\item the ECALp, electromagnetic calorimeters for positrons (based on the HighCompactCalo concept).
\end{itemize}

Importantly, ECALp is not only a detector for LUXE. Its compact geometry, sensor technology and integration philosophy serve as prototypes for forward luminometers proposed for FCC-ee, ILC and LCF. The proponents of these luminometer concepts and of ECALp largely overlap, ensuring technological continuity between LUXE and future Higgs Factories.

At present, both SiWECAL and HighCompactCalo/ECALp employ the same type of highly granular silicon pad sensors~\cite{Tomita:2014wha}. This convergence provides a strong rationale for establishing a unified infrastructure capable of supporting both detector concepts.

The \textbf{TARDIS Lab\footnote{
Its name is TARDIS Lab~\cite{DoctorWho} because it \textit{looks bigger on the inside.}}} has been established at IFIC to provide such an infrastructure, enabling silicon sensor characterization and ultra-thin module assembly for both SiWECAL and ECALp developments.

For \textbf{SiWECAL}, the main technological challenge is the integration of large-area silicon sensors onto thin PCB-based readout layers while preserving planarity, mechanical stability and reliable electrical contact over the full sensor surface. Deformation of the PCB and mechanical stress at the sensor--PCB interface can lead to delamination or loss of electrical connectivity. At IFIC, these effects were mitigated through a hybrid integration technique combining CNC-cut 3M double-sided tape for mechanical fixation with localized conductive-epoxy contacts for electrical interconnection, thereby decoupling mechanical strain from the electrical bonds.

For \textbf{ECALp}, the primary challenge is extreme compactness. At IFIC, assembly procedures have been developed to guarantee that the full active module stack (mechanical support, fan-out, sensor and high-voltage layer) remains below 1~mm thickness, preserving the compact geometry required for forward calorimetry and precision luminometry.

The TARDIS Lab provides a common technological platform supporting both highly integrated barrel calorimetry and ultra-compact forward calorimetry within DRD Calo, the FCAL Collaboration and LUXE.


\section{TARDIS Lab Facility}
\label{sec:infra}

\begin{figure*}[t]
\centering
\includegraphics[width=\textwidth]{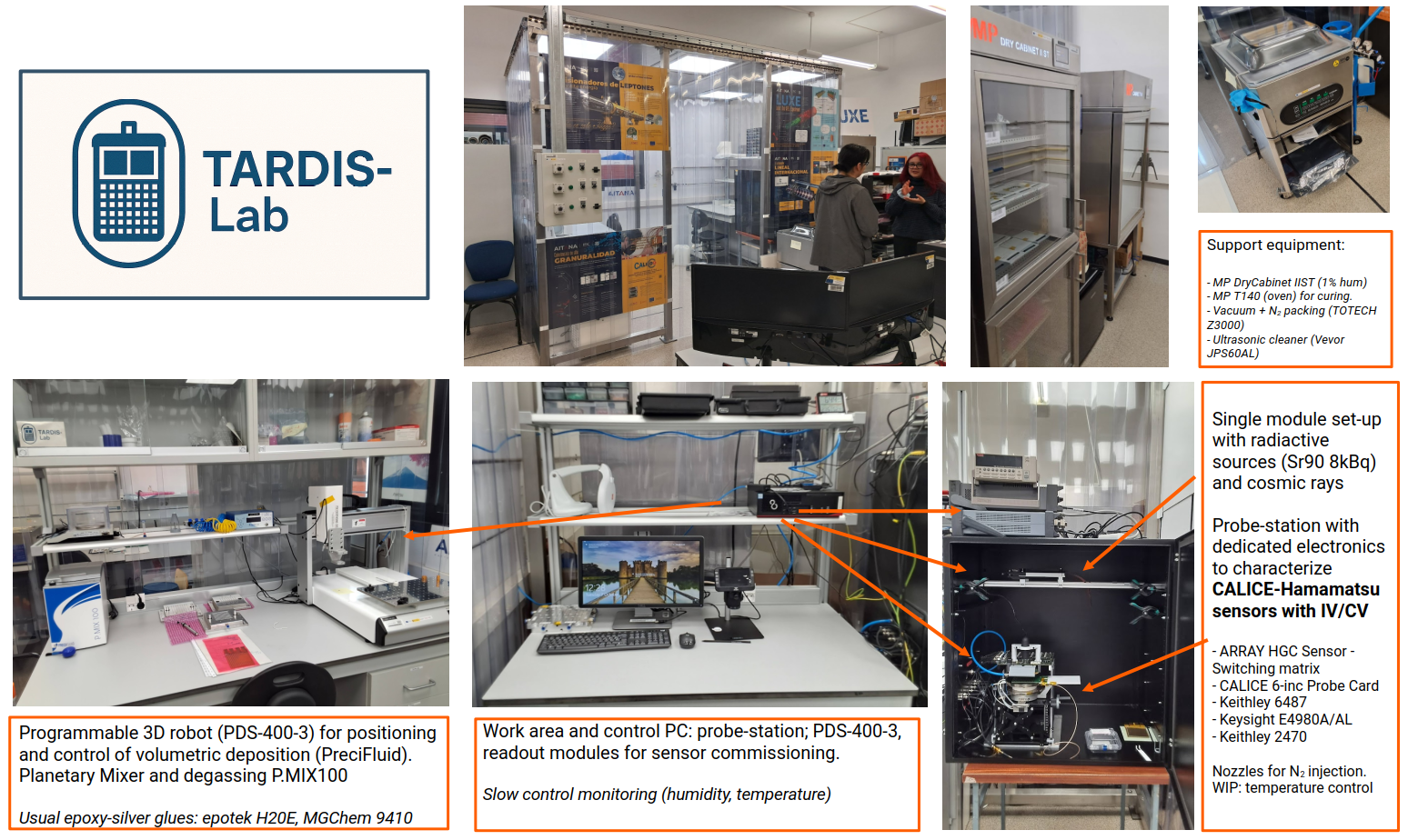}
\caption{Overview of the TARDIS Lab facility at IFIC. Top left: TARDIS Lab logo. Top center/right: ISO7 clean-room area and controlled storage (dry cabinet). 
Bottom left: programmable 3D robotic positioning system (PDS-400-3) used together with a Precifluid volumetric dispenser for controlled adhesive deposition. Bottom center: control and monitoring workstation. Bottom right: probe-station setup with dedicated electronics for pad-by-pad IV/CV characterization of CALICE–Hamamatsu silicon sensors.}
\label{fig:tardis_overview}
\end{figure*}

The TARDIS Lab has been conceived as a dedicated infrastructure for silicon sensor qualification and ultra-thin calorimeter module assembly, serving both SiWECAL and HighCompactCalo/ECALp developments. It provides a controlled clean-room environment together with specialized equipment for precision assembly and systematic electrical characterization.

\subsection{Clean-Room Environment and Controlled Storage}

The facility operates within a semi-portable ISO7 clean-room environment with local ISO5 conditions for critical operations. Silicon sensors and front-end boards are stored in humidity-controlled cabinets and undergo controlled drying before assembly to 
minimize residual oxidation of conductive layers and PCB deformation.

\subsection{Precision Assembly Capabilities}
Ultra-thin module assembly requires micrometric control of mechanical alignment and adhesive deposition. 
The main elements of the assembly chain are shown in Fig.~\ref{fig:tardis_overview}, including the programmable robotic positioning system (PDS) combined with a volumetric dispensing unit (Precifluid) for controlled adhesive deposition, together with precision alignment tooling.
The assembly workflow follows documented laboratory procedures for adhesive preparation, deposition and quality control. Standardized protocols for conductive epoxy mixing and module assembly are maintained together with sensor-level quality-control documentation for each device, ensuring reproducibility and full component traceability throughout the production chain~\cite{TARDISGluePreparation,TARDISAssemblyProcedure,TARDISSensorQC}.

A two-component electrically conductive epoxy (EPO-TEK H20E~\cite{EPO_H20E}) is currently used as the baseline solution for electrical interconnection in both SiWECAL (ECALe) and ECALp modules. In parallel, alternative commercial one- and two-component conductive epoxies~\cite{MG8331S,MG9410} requiring lower curing temperatures are being evaluated. In particular, one-component formulations eliminate operator-dependent variations during mixing and may therefore simplify handling and improve process reproducibility.

\begin{figure*}[ht!]
\centering
\includegraphics[width=\textwidth]{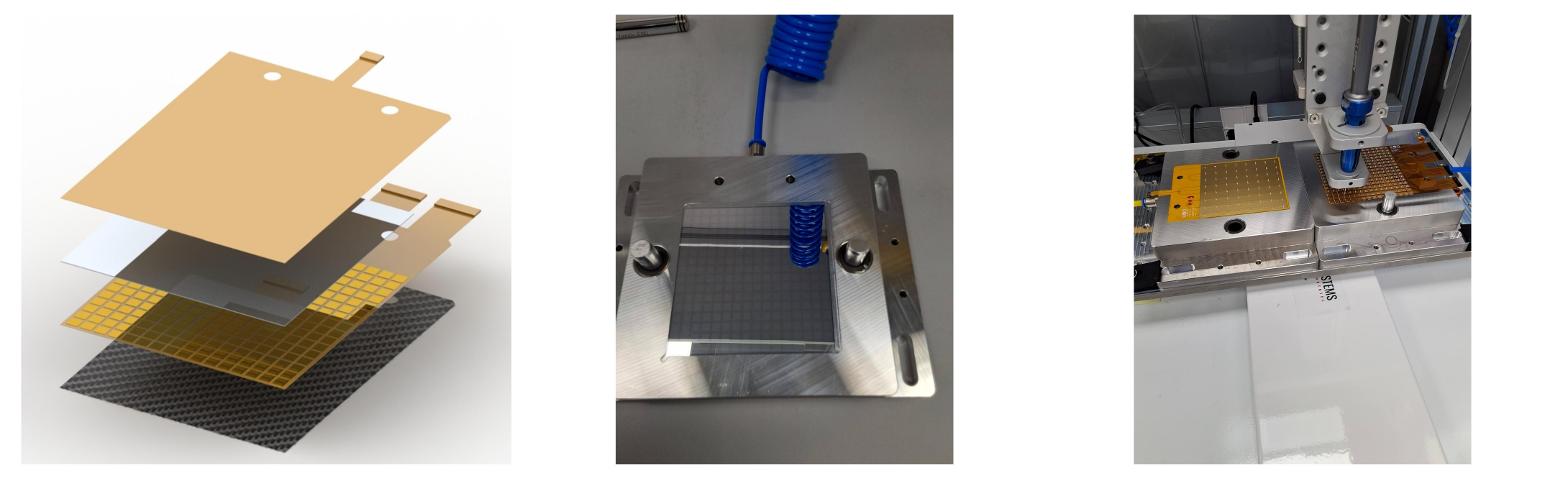}
\caption{
ECALp module architecture and dedicated assembly tooling used at the TARDIS Lab.
Left: exploded view of the compact sensor interface system (CSIS), comprising the mechanical support, silicon sensor, fan-out and high-voltage layers.
Center: silicon sensor positioned in its dedicated alignment jig prior to assembly.
Right: high-voltage and fan-out layers held in their respective positioning jigs during controlled adhesive deposition.
The dedicated tooling provides mechanical referencing and reproducible relative positioning of the different CSIS components while preserving the compactness of the assembled module.
}
\label{fig:ecalp_assembly}
\end{figure*}

The ECALp module architecture and the main stages of its assembly are illustrated in Fig.~\ref{fig:ecalp_assembly}. Dedicated positioning jigs are used for the silicon sensor, fan-out and high-voltage layers, providing mechanical referencing and reproducible relative alignment during adhesive deposition and integration.
A similar assembly philosophy and dedicated tooling are used for SiWECAL modules, adapted to their different PCB-based architecture and mechanical constraints.

For ECALp modules, where extreme compactness is the primary constraint, adhesive dots are typically deposited with diameters of approximately 3~mm and thicknesses below 50~\textmu m. The thickness is controlled to remain above the critical conductivity threshold (typically 10--20~\textmu m for silver-loaded conductive epoxies) while minimizing the overall module stack height. This controlled deposition ensures reliable electrical contact without compromising the sub-millimeter compactness requirement.

For SiWECAL (ECALe) modules, mechanical stability is the dominant constraint due to the integration with PCBs. A hybrid bonding solution has been implemented to resolve sensor delamination issues previously observed as a consequence of PCB deformation. Mechanical stabilization is provided by 3M double-sided adhesive tape (VHB 5907F), with a nominal thickness of 200~\textmu m, cut in stencil form to ensure controlled placement and stress decoupling. The conductive epoxy is then used to provide the electrical connection. In this configuration, the effective bonding thickness is significantly larger than in ECALp modules, allowing mechanical strain to be absorbed by the adhesive layer while preserving electrical reliability.

This differentiated bonding strategy allows the TARDIS Lab to address the distinct mechanical constraints of highly integrated barrel calorimetry (SiWECAL) and ultra-compact forward calorimetry (ECALp).

\subsection{Electrical Characterization Capabilities}

\begin{figure}[ht!]
\centering
\includegraphics[width=0.6\textwidth]{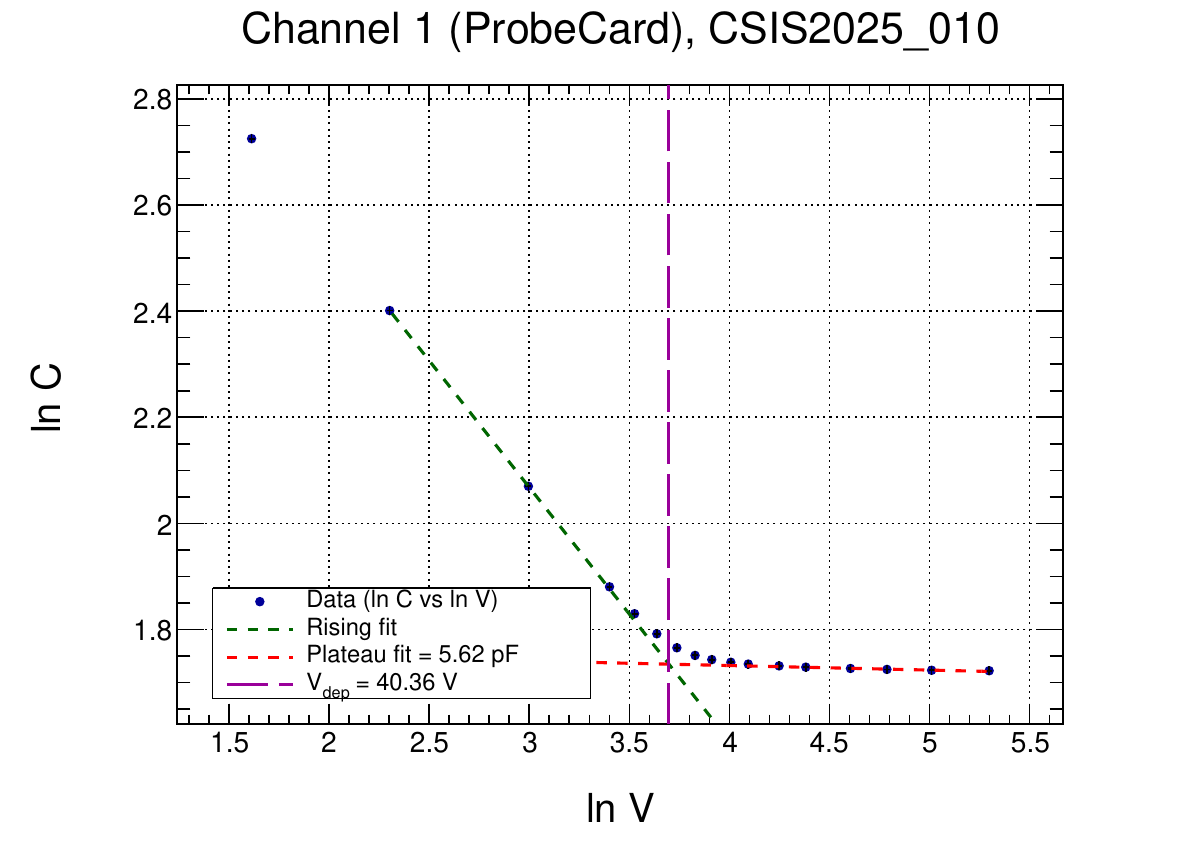}
\caption{
Example of a capacitance--voltage (C--V) measurement performed on a representative pad after assembly at CSIS level. The measurement was carried out at IFIC using the CALICE sensor probe card and an intermediate PCB connecting the probe-card pogo pins to the CSIS fan-out. The depletion voltage is obtained from the intersection of linear fits to the partially depleted and fully depleted regions in the $\ln C$--$\ln V$ representation.
}
\label{fig:csis_cv}
\end{figure}

Sensor qualification is performed using a probe-station~\cite{SensorTestCALICE} setup with custom probe cards (CERN–FCAL–AIDAInnova), high-accuracy SMUs, picoammeters and an LCR meter, enabling pad-by-pad IV and CV characterization of the PIN sensors. Data are stored in structured form, with automated database integration foreseen in the next upgrade phase.
The same probe-card system used for bare-sensor characterization is also employed for assembled compact sensor interface system (CSIS) modules through an intermediate PCB connecting the probe-card pogo pins to the CSIS fan-out. This configuration enables pad-by-pad IV and CV measurements after assembly and provides a direct verification of the electrical behaviour and connectivity of the integrated modules.
Figure~\ref{fig:csis_cv} shows a representative C--V measurement performed after assembly at CSIS level. The depletion voltage is obtained from the intersection of linear fits to the partially depleted and fully depleted regions in the log$C$--log$V$ representation, demonstrating that the integrated modules can be electrically characterized pad by pad after assembly.

\subsection{Unified Support for SiWECAL and HighCompactCalo}

A key feature of the TARDIS Lab is its ability to support both SiWECAL and HighCompactCalo developments within a common framework. Both detector lines currently employ the same type of highly granular silicon pad sensors, enabling shared qualification workflows and assembly procedures.

This unified infrastructure ensures coherent sensor characterization standards, scalable module production capability, cross-fertilization between barrel and forward calorimetry R\&D, and efficient transition from DRD instrumentation developments to experiment-oriented detector realization within the FCAL Collaboration.

It is important to emphasize that the developments described here are part of a broader international effort. Silicon sensor production, front-end electronics design, mechanical structures, simulation studies and beam-test campaigns are carried out by multiple institutes within the DRD Calo and FCAL Collaboration frameworks. The TARDIS Lab does not replace these distributed contributions; rather, it provides a dedicated assembly and qualification hub where components developed by international partners are integrated, validated and prepared for system-level operation. In this sense, the TARDIS Lab acts as a unifying interface — effectively the “glue” — between sensor development, electronics R\&D and full detector realization.


\section{Module Production, Integration and Test-Beam Validation}
\label{sec:production}

The assembly and characterization procedures described above have been applied to the production and validation of SiWECAL and ECALp modules.

\subsection{SiWECAL Demonstrators}

Two SiWECAL demonstrator modules were fully assembled at IFIC, including sensor qualification, hybrid mechanical--electrical bonding and post-assembly validation.

Sensor-to-PCB connectivity was evaluated using the analogue-probe functionality of the SKIROC2A ASICs implemented in the FEV2.1 active sensor units. For each channel, the analogue preamplifier voltage was measured with the sensor high voltage connected and disconnected. A properly connected silicon pad produces the expected shift in the preamplifier operating point when the sensor bias is applied, whereas the absence of this shift indicates an open or defective connection between the sensor pad and the PCB input. Applied to the two modules assembled in 2025, this method identified connectivity problems in approximately $10^{-3}$ of the instrumented pads.

The modules were subsequently tested with electron beams at DESY in a configuration without tungsten absorbers ($0~X_0$). No mechanical failures or sensor delamination were observed after assembly, transport and beam operation, and both modules remained fully operational throughout the campaign. Laboratory connectivity measurements repeated in 2026 on the same modules yielded consistent results, supporting the robustness of the hybrid 3M VHB 5907F and conductive-epoxy bonding approach~\cite{2025SiWECALTB}.

\subsection{ECALp Multi-Layer Validation}

For ECALp, 20 ultra-compact modules were assembled and tested at DESY in configurations ranging from $0~X_0$ to $20~X_0$, enabling studies of electromagnetic shower development, containment, linearity and mechanical stability of the sub-millimetre integration concept.

The thickness of the assembled CSIS modules was measured after integration. All 20 assembled CSIS modules were measured to have a thickness below 800~\textmu m, with an average value of approximately 760~\textmu m, comfortably below the design target of 1--1.2~mm.

Seventeen of the 20 modules, corresponding to all units currently available at IFIC, were subsequently tested channel by channel in the laboratory. Approximately 0.5\% of the $17\times256=4352$ tested channels showed connectivity problems. A preliminary comparison with the corresponding test-beam data indicates a compatible fraction of non-responsive or anomalous channels.

Together, the measured thickness and the low fraction of connectivity problems demonstrate that the ECALp assembly procedures meet the principal mechanical and electrical requirements of the compact module concept. Results for individual ECALp sensor modules are reported in Ref.~\cite{Irles_2026}.

\subsection{From Prototype to Structured Production}

The assembly and successful beam operation of the SiWECAL and ECALp modules demonstrate the viability of both integration concepts. For ECALp, the measured average CSIS thickness of approximately 760~\textmu m and the 0.5\% fraction of channels showing connectivity problems confirm that the principal mechanical and electrical requirements of the compact module concept have been achieved. For the two SiWECAL modules assembled in 2025, connectivity problems were identified at the level of approximately $10^{-3}$ of the instrumented pads, while no mechanical failures or sensor delamination were observed after assembly, transport and beam operation. These results provide the quantitative basis for the transition toward full tower and calorimeter integration foreseen in the 2026--2027 roadmap.


\section{Prospects 2026--2027}
\label{sec:prospects}

The next phase of TARDIS Lab activities follows milestones aligned with the DRD Calo roadmap and the FCAL Collaboration detector programs.

\subsection{2026: Completion of a Full SiWECAL Tower}
In 2026, the objective is the completion and systematic validation of a 15-module SiWECAL calorimeter tower, including studies of multi-layer mechanical tolerances, electrical integration and response uniformity.

\subsection{2027: Completion of the Full ECALp Calorimeter}

For ECALp, the 2027 objective is the completion of the full calorimeter assembly based on the ultra-compact module concept.

The program includes structured production of ultra-compact modules, integration of next-generation sensors, optimization of interconnection schemes and full calorimeter assembly with absorber structure, followed by system-level validation.

This milestone consolidates ECALp not only as a detector for LUXE but also as a technological demonstrator for forward luminometers at future Higgs Factories.

\subsection{Upgrades to the TARDIS Lab}

Several infrastructure upgrades are being implemented to support the transition from prototype development to structured module production.

The robotic dispensing infrastructure combines the programmable 3D robotic positioning system (PDS) with a Precifluid volumetric dispenser used for controlled deposition of conductive adhesives. While the robotic positioning system remains unchanged, upgraded tooling has recently been implemented for the Precifluid dispenser, allowing larger adhesive cartridges to be used during operation. This modification increases the dispensing capacity per iteration from 3~mL to 5~mL, reducing interruptions during assembly sequences and improving operational efficiency.

In parallel, a new iteration of dedicated assembly fixtures is being produced to increase the module handling capacity. The updated tooling is designed to approximately triple the number of modules that can be processed simultaneously during key stages of the assembly workflow, enabling higher throughput in preparation for structured production campaigns.

Additional upgrades include active Peltier-based temperature regulation with closed-loop control inside the probe-station Faraday cage, together with continuous monitoring of temperature and humidity. These developments will provide controlled and documented environmental conditions during sensor and CSIS characterization. 

Together, these upgrades represent an important step in the evolution of the TARDIS Lab from a prototype-development environment toward a facility capable of supporting scalable detector module production.


\section{Conclusions}
\label{sec:conclusions}

The TARDIS Lab at IFIC provides dedicated infrastructure for the characterization and assembly of highly granular silicon calorimeter modules, supporting both SiWECAL within DRD Calo and the ultra-compact ECALp concept within the FCAL Collaboration. The dimensional and electrical characterization of the ECALp modules confirms that the assembly procedure meets the principal requirements of the compact sensor-layer concept. For SiWECAL, analogue-probe measurements, repeated laboratory tests and successful beam operation support the mechanical and electrical robustness of the hybrid bonding approach. Together with the DESY test-beam campaigns, these results demonstrate the reliability of the implemented assembly and qualification workflows and their suitability for larger-scale detector integration. The planned activities for 2026--2027 will extend this work toward complete SiWECAL tower and ECALp calorimeter systems for LUXE and future Higgs Factories.


\section*{Acknowledgements}
\addcontentsline{toc}{section}{Acknowledgements}
We acknowledge the financial support
from: the Spanish MICIU/AEI and European Union/FEDER via the grant \texttt{PID2021-122134NB-C21} the Generalitat Valenciana (GV) via the Excellence Grant \texttt{CIPROM/2021/073} and the PlanGenT program with the grant number \texttt{CIDEGENT/2020/021}; the MCIN with funding from the European Union NextGenerationEU and Generalitat Valenciana in the call ``Programa de Planes Complementarios de I+D+i (PRTR 2022)'' through the project with reference \texttt{ASFAE/2022/015}; and the Program Programa Estatal
para Desarrollar, Atraer y Retener Talento PEICTI 2021-2023 through the project
with reference \texttt{CNS2022-135420}.

\printbibliography

@article{Behnke:2013xla,
    editor = "Behnke, Ties and Brau, James E. and Foster, Brian and Fuster, Juan and Harrison, Mike and Paterson, James McEwan and Peskin, Michael and Stanitzki, Marcel and Walker, Nicholas and Yamamoto, Hitoshi",
    title = "{The International Linear Collider Technical Design Report - Volume 1: Executive Summary}",
    eprint = "1306.6327",
    archivePrefix = "arXiv",
    primaryClass = "physics.acc-ph",
    reportNumber = "ILC-REPORT-2013-040, ANL-HEP-TR-13-20, BNL-100603-2013-IR, IRFU-13-59, CERN-ATS-2013-037, COCKCROFT-13-10, CLNS-13-2085, DESY-13-062, FERMILAB-TM-2554, IHEP-AC-ILC-2013-001, INFN-13-04-LNF, JAI-2013-001, JINR-E9-2013-35, JLAB-R-2013-01, KEK-REPORT-2013-1, KNU-CHEP-ILC-2013-1, LLNL-TR-635539, SLAC-R-1004, ILC-HIGRADE-REPORT-2013-003",
    month = "6",
    year = "2013"
}

@article{FCC:2025lpp,
    author = "Benedikt, M. and others",
    collaboration = "FCC",
    title = "{Future Circular Collider Feasibility Study Report: Volume 1, Physics, Experiments, Detectors}",
    eprint = "2505.00272",
    archivePrefix = "arXiv",
    primaryClass = "hep-ex",
    reportNumber = "CERN-FCC-PHYS-2025-0002",
    doi = "10.1140/epjc/s10052-025-15077-x",
    journal = "Eur. Phys. J. C",
    volume = "85",
    number = "12",
    pages = "1468",
    year = "2025"
}

@article{LinearCollider:2025lya,
    author = "Abramowicz, H. and others",
    collaboration = "Linear Collider",
    title = "{The Linear Collider Facility (LCF) at CERN}",
    eprint = "2503.24049",
    archivePrefix = "arXiv",
    primaryClass = "hep-ex",
    reportNumber = "DESY-25-054, FERMILAB-PUB-25-0239-CSAID",
    month = "3",
    year = "2025"
}

@article{CEPCStudyGroup:2023quu,
    author = "Abdallah, Waleed and others",
    collaboration = "CEPC Study Group",
    title = "{CEPC Technical Design Report: Accelerator}",
    eprint = "2312.14363",
    archivePrefix = "arXiv",
    primaryClass = "physics.acc-ph",
    reportNumber = "IHEP-CEPC-DR-2023-01, IHEP-AC-2023-01",
    doi = "10.1007/s41605-024-00463-y",
    journal = "Radiat. Detect. Technol. Methods",
    volume = "8",
    number = "1",
    pages = "1--1105",
    year = "2024",
    note = "[Erratum: Radiat.Detect.Technol.Methods 9, 184--192 (2025)]"
}

@article{Thomson:2009rp,
    author = "Thomson, M. A.",
    title = "{Particle Flow Calorimetry and the PandoraPFA Algorithm}",
    eprint = "0907.3577",
    archivePrefix = "arXiv",
    primaryClass = "physics.ins-det",
    reportNumber = "CU-HEP-09-11",
    doi = "10.1016/j.nima.2009.09.009",
    journal = "Nucl. Instrum. Meth. A",
    volume = "611",
    pages = "25--40",
    year = "2009"
}

@article{Sefkow:2015hna,
    author = {Sefkow, Felix and White, Andy and Kawagoe, Kiyotomo and P{\"o}schl, Roman and Repond, Jos{\'e}},
    title = "{Experimental Tests of Particle Flow Calorimetry}",
    eprint = "1507.05893",
    archivePrefix = "arXiv",
    primaryClass = "physics.ins-det",
    reportNumber = "DESY-14-032, KYUSHU-RCAPP-2015-01, LAL-15-235",
    doi = "10.1103/RevModPhys.88.015003",
    journal = "Rev. Mod. Phys.",
    volume = "88",
    pages = "015003",
    year = "2016"
}

@misc{CALICE,
  author = {{CALICE Collaboration}},
  title  = {CALICE Collaboration website},
  year   = {2026},
  url    = {https://twiki.cern.ch/twiki/bin/view/CALICE/WebHome}
}

@misc{FCAL,
  author = {{FCAL Collaboration}},
  title  = {FCAL Collaboration website},
  year   = {2026},
  url    = {https://fcal.desy.de}
}

@misc{DRDCalo,
  author = {{CERN DRD Calorimetry Collaboration}},
  title  = {Detector R\&D Collaboration for Calorimetry (DRD Calo)},
  year   = {2024},
  url    = {https://calorimetry.web.cern.ch/drd}
}

@article{Kawagoe:2019dzh,
    author = "Kawagoe, K. and others",
    title = "{Beam test performance of the highly granular SiW-ECAL technological prototype for the ILC}",
    eprint = "1902.00110",
    archivePrefix = "arXiv",
    primaryClass = "physics.ins-det",
    reportNumber = "KYUSHU-RCAPP-2019-04, AIDA-2020-PUB-2020-001",
    doi = "10.1016/j.nima.2019.162969",
    journal = "Nucl. Instrum. Meth. A",
    volume = "950",
    pages = "162969",
    year = "2020"
}

@article{LumiCal,
  author  = {H. Abramowicz and others},
  title   = {Forward instrumentation for ILC detectors},
  journal = {JINST},
  volume  = {5},
  year    = {2010},
  pages   = {P12002},
  doi     = {10.1088/1748-0221/5/12/P12002}
}

@article{BeamCal,
  author  = {H. Abramowicz and others},
  title   = {Instrumentation of the very forward region of the ILC detectors},
  journal = {IEEE Transactions on Nuclear Science},
  volume  = {57},
  year    = {2010},
  pages   = {425},
  doi     = {10.1109/TNS.2009.2035464}
}

@inproceedings{Tomita:2014wha,
    author = "Tomita, Tatsuhiko and others",
    title = "{A study of silicon sensor for ILD ECAL}",
    booktitle = "{International Workshop on Future Linear Colliders}",
    eprint = "1403.7953",
    archivePrefix = "arXiv",
    primaryClass = "physics.ins-det",
    month = "3",
    year = "2014"
}

@misc{DoctorWho,
  title        = {Doctor Who},
  author       = {{BBC}},
  year         = {1963},
  howpublished = {British Broadcasting Corporation television series},
  note         = {First broadcast 23 November 1963}
}

@misc{EPO_H20E,
  author       = {{Epoxy Technology Inc.}},
  title        = {EPO-TEK H20E Electrically Conductive Epoxy},
  year         = {2024},
  howpublished = {Technical datasheet},
  url          = {https://meridianadhesives.com/products/epo-tek-h20e/},
  note         = {Epoxy Technology Inc., Billerica, Massachusetts, USA}
}

@misc{SensorTestCALICE,
  author = {M. Almanza-Soto},
  title  = {SensorTest\_CALICE\_6in\_256ch: Software and documentation for silicon sensor characterization},
  year   = {2024},
  howpublished = {\url{https://github.com/almanzam218/SensorTest_CALICE_6in_256ch/wiki/README}},
  note   = {GitHub wiki, accessed 2026-03-06}
}

@misc{MG8331S,
  author       = {{MG Chemicals}},
  title        = {8331S Silver Conductive Epoxy Adhesive Technical Datasheet},
  year         = {2024},
  url          = {https://mgchemicals.com/products/adhesives/electrically-conductive-adhesives/silver-conductive-epoxy-8331s/},
  note         = {MG Chemicals, Burlington, Ontario, Canada}
}

@misc{MG9410,
  author       = {{MG Chemicals}},
  title        = {9410 One-Part Conductive Epoxy Adhesive Technical Datasheet},
  year         = {2024},
  url          = {https://mgchemicals.com/products/adhesives/electrically-conductive-adhesives/9410-one-part-conductive-epoxy/},
  note         = {MG Chemicals, Burlington, Ontario, Canada}
}

@misc{2025SiWECALTB,
  author       = {V. Boudry},
  title        = {SiW-ECAL Status and Test-Beam Results},
  year         = {2025},
  howpublished = {Talk at the CERN DRD6 Collaboration Meeting},
  url          = {https://indico.ijclab.in2p3.fr/event/11400/contributions/37689/attachments/25380/37307/2025-03-25@DRD6_SiW-ECAL-status_VB.pdf},
  note         = {IJCLab DRD6 meeting, 25 March 2025}
}

@misc{TARDISGluePreparation,
  author       = {{TARDIS Lab, IFIC}},
  title        = {Conductive Epoxy Preparation and Mixing Procedure for Silicon Calorimeter Module Assembly},
  year         = {2025},
  url          = {https://cernbox.cern.ch/s/muLT1SG7XIzH3hj},
  note         = {TARDIS Lab internal technical documentation}
}

@misc{TARDISAssemblyProcedure,
  author       = {{TARDIS Lab, IFIC}},
  title        = {Ultra-thin Silicon Calorimeter Module Assembly Procedure},
  year         = {2025},
  url          = {https://cernbox.cern.ch/s/fAhtPN8vnCbhniS},
  note         = {TARDIS Lab internal technical documentation}
}

@misc{TARDISSensorQC,
  author       = {{TARDIS Lab, IFIC}},
  title        = {Sensor Quality-Control Datasheet Template for Calorimeter Module Assembly},
  year         = {2025},
  url          = {https://cernbox.cern.ch/s/uo46DtlCAoUYMsN},
  note         = {TARDIS Lab quality control documentation}
}

@article{Irles_2026,
   title={Test of a partly instrumented highly compact and granular electromagnetic calorimeter in an electron beam of 1–6 GeV},
   volume={141},
   ISSN={2190-5444},
   url={http://dx.doi.org/10.1140/epjp/s13360-026-07926-9},
   DOI={10.1140/epjp/s13360-026-07926-9},
   number={6},
   journal={The European Physical Journal Plus},
   publisher={Springer Science and Business Media LLC},
   author={Irles, Adrián and },
   year={2026},
   month=June }

\end{document}